\documentclass[showpacs,pre]{revtex4}
\makeatletter
\def\place@bibnumber@inl#1{#1.}%

\makeatother

\usepackage{here}
\usepackage{epsf}
\usepackage{amsmath}
\usepackage{times,mathmine}
\begin{document}
\title[Leaf]{The Shape of the Edge of a Leaf}
\author{M. Marder}
\affiliation{Center for Nonlinear Dynamics and Department of Physics\\
The University of Texas at Austin, Austin TX 78712\\
marder@chaos.ph.utexas.edu}

\begin{abstract}
  Leaves and flowers frequently have a characteristic rippling pattern
  at their edges.  Recent experiments found similar patterns in torn
  plastic. These patterns can be reproduced by imposing metrics upon
  thin sheets. The goal of this paper is to discuss a collection of
  analytical and numerical results for the shape of a sheet with a
  non--flat metric. First, a simple condition is found to determine
  when a stretched sheet folded into a cylinder loses axial symmetry,
  and buckles like a flower. General expressions are next found for
  the energy of stretched sheet, both in forms suitable for numerical
  investigation, and for analytical studies in the continuum. The
  bulk of the paper focuses upon long thin strips of material
  with a linear gradient in metric. In some special cases, the
  energy--minimizing shapes of such strips can be determined
  analytically. Euler--Lagrange equations are found which determine
  the shapes in general. The paper closes with numerical
  investigations of these equations.
\end{abstract}
\maketitle

\section{Introduction}

\begin{figure}[H]
\begin{center}
\epsfysize2.5in\epsffile{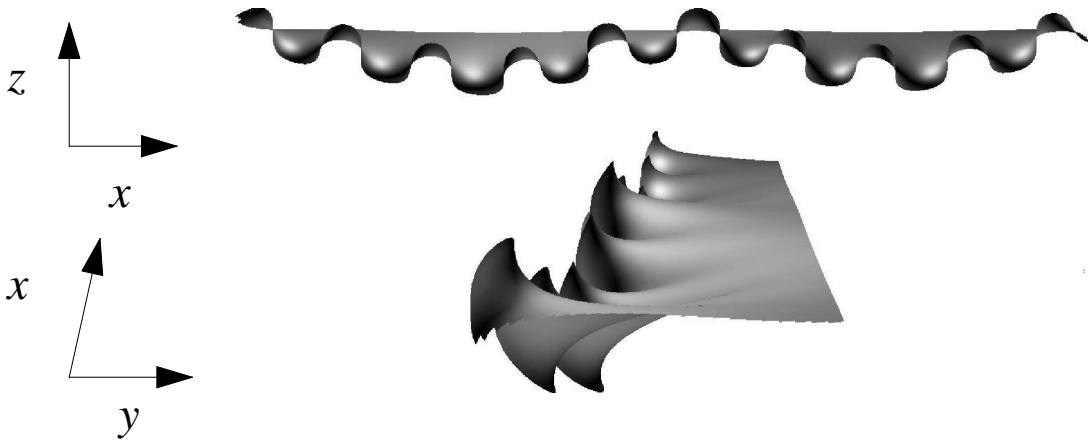}
\caption{\small Illustration of shapes produced by stretching a membrane at
  one end. This figure was produced by numerical minimization of
  Eq.~\ref{eq:energy2} with $\sqrt{g_{xx}(y)}=
  .7\exp(-y/12)+.3\exp(-y/36)$.  The numerical sample was
  600 units long, 60 units wide, and two layers high.}
\label{fig:Membrane_setup}
\end{center}
\end{figure}

\def\mermintro{ In choosing a topic for this article in honor of David
  Mermin, I must confess that I have settled upon something odd. The
  subject is a rippled pattern that often appears at the edges of
  leaves and flowers.  One can even produce it by ripping in half a
  thin sheet of plastic, such as a plastic bag\cite{Sharon.02}. A
  numerical example of a stretched sheet with two generations of waves
  appears in Figure \ref{fig:Membrane_setup}. The problem and its
  solution, like many of Mermin's writings, has a whimsical twist.
  
  Maybe this is not a subject that would normally be regarded as
  fundamental, but it does touch on an interesting basic question,
  which concerns the complexity of the instructions needed to generate
  complex natural patterns. The dominant belief among biologists is
  that the curling shapes of plants are produced by detailed genetic
  instructions telling various sections to curl up and down\cite{Byrne.01}. The
  physics community interested in studying patterns would prefer to
  assume that complex forms are produced when possible by simple
  rules, a view with some support also among
  biologists\cite{Green.96}. It is natural to guess that by imposing
  non--flat metrics 
  upon thin sheets, elasticity alone will compel them to curl
  spontaneously into fractal forms. However, to proceed from a guess
  to a detailed demonstration requires some effort. 

}
  
\def\nonmermintro{ A characteristic rippled pattern often appears at
  the edges of leaves and flowers.  One can even produce it by ripping
  in half a thin sheet of plastic, such as a plastic
  bag\cite{Sharon.02}. A numerical example of a stretched sheet with
  two generations of waves appears in Figure \ref{fig:Membrane_setup}.
  I have found a case where the mathematics of this problem can be
  solved exactly.
  
  This topic touches on an interesting basic question, which concerns
  the complexity of the instructions needed to generate complex
  natural patterns. The dominant belief among biologists is that the
  curling shapes of plants are produced by detailed genetic
  instructions telling various sections to curl up and
  down\cite{Byrne.01}. The physics community interested in studying
  patterns would prefer to assume that complex forms are produced when
  possible by simple rules, a view with some support also among
  biologists\cite{Green.96,Green.99}. It is natural to guess that by imposing
  non--flat metrics upon thin sheets, elasticity alone will compel
  them to curl spontaneously into fractal forms. However, to proceed
  from a guess to a detailed demonstration requires some effort.  }

\nonmermintro

The basic question addressed throughout this paper is the following:
Suppose one has a rectangular sheet of material, described by
coordinates $x$ and $y$. Take this rectangular strip and impose a new
metric on it so that distances $dr$ between nearby points originally
separated by $(dx,dy)$ are given by
\begin{equation}
dr^2=g_{xx}dx^2+2g_{xy}dxdy+g_{yy}dy^2.
\end{equation}
Embed the sheet in three--dimensional space, allowing it to curl
as needed to obey this new metric. What shape does it take?

The main results in the paper are the following:
\begin{itemize}
\item Use of results from differential geometry to find when
  axisymmetric sheets wrapped into cylinders are stable against
  buckling (Section II).
\item Energy functional for a model of a thin sheet with non--flat
  metric as a collection of interacting mass points (Section III).
\item Derivation of continuum nonlinear elastic theory from the
  original energy functional (Section IV)
\item Definition of solvable problem in thin strip with linear gradient in
  metric in terms of a set of ordinary differential equations (Section V).
\item Exact solution of these ordinary differential equations in terms
  of elementary function when certain bending and torsion angles in
  them are specified in specific ways (Sections VI--VIII).
\item Calculation of continuum energy functional in terms of the
  solutions of the ordinary differential equations (Section IX). 
\item Derivation of Euler--Lagrange equations from continuum energy
  functional, demonstration that analytical solutions of previous
  sections correspond to energy minimizing solutions when special
  boundary conditions are imposed (Section X).
\item Numerical solutions of Euler--Lagrange equations to find shapes
  of strips for more general boundary conditions (Section XI).
\end{itemize}

Nechaev and Voituriez\cite{Nechaev.01} and Henderson and Taimina 
\cite{Taimina.02} have carried out the only other studies along these
lines of which I am aware. The first group of authors provides an exact
solution using conformal mapping techniques to analyze an
exponentially growing metric, while the second set of authors provides
additional information on the mathematics of hyperbolic planes. Both
studies are nicely complementary to the work presented here.

\section{Buckling of Flowers}

 \begin{figure}[!ht]
(a) \epsfxsize=3in\epsffile{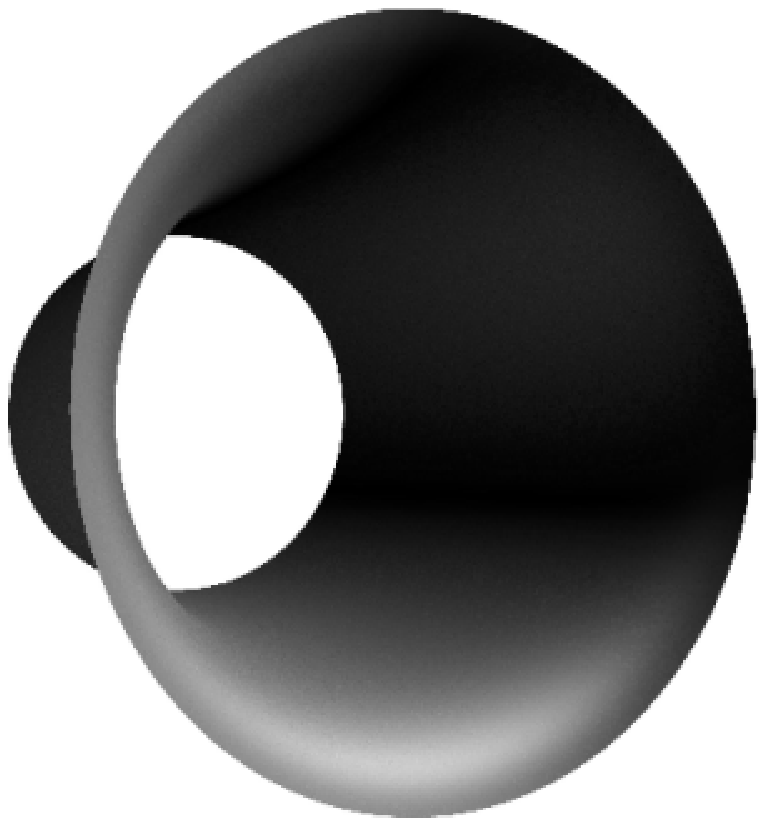}\\
(b)\epsfxsize=3in\epsffile{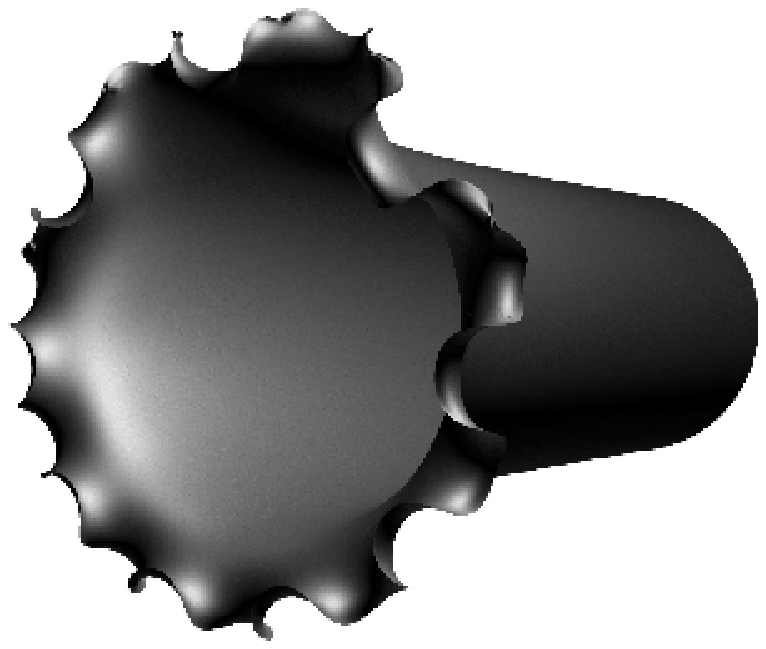}
 \caption{Result of numerical minimization of Eq.~\ref{eq:energy2}
   for system twisted around into a cylinder. The $x$ coordinate of
   the original material travels in an angular direction around the
   axis, while the $y$ coordinate of the original material describes
   motion along the axis. (a) The metric is $\sqrt{g_{xx}}=1/(1+2\pi
   y/300)$, and the original radius of the cylinder is
   $R=47.8=300/(2\pi)$. The product $R_\infty\kappa$ in
   Eq.~\ref{eq:G-B6} equals 1. The edge is without ripples and has
   splayed out to lie in a plane perpendicular to the axis. (b) Same
   as (a) except that the metric is $\sqrt{g_{xx}}=1/(1+y/2)$. The
   bound in Eq.~\ref{eq:G-B6} is exceeded by a factor of 25. There are
   around 22 ripples at the edge. }
 \label{fig:trumpet}
 \end{figure}

It would be very appealing if one could employ differential geometry alone to
resolve questions about the buckling of sheets with non--flat metrics.
I have not had success along these lines for the problem depicted in
Fig.~\ref{fig:Membrane_setup}. However, if I digress briefly to
discuss flowers rather than leaves, I can make some progress, and want
to mention this result before passing on to the main topic of the
paper.

Suppose that one adds an additional constraint to the problem, which
is that when the sheet is embedded in three dimensions, it is required
to be periodic along $x$, with period $2\pi R_\infty$. One has a
distorted cylinder, as shown in Figs.~\ref{fig:trumpet}.  One can use results from differential geometry to
obtain a simple criterion for when such a structure loses its
cylindrical symmetry and buckles. The Gauss--Bonnet
theorem\cite{Pogorelov.56} applied to this case says that
\begin{equation}
\int ds\,\kappa=-\int dA\, K,
\label{eq:G-B}
\end{equation}
where $\kappa$ is the geodesic curvature, and $K$ is the Gaussian
curvature. The first integral is a line integral taken around the edge
of the cylinder, while the second is a surface integral.

Use $x$ and $y$ to refer to material coordinates. For a metric where only
$g_{xx}(y)$ differs from the value expected in flat space, the
Gauss--Codazzi relations\cite{Pogorelov.56} give
\begin{equation}
K=-{1\over \sqrt{g_{xx}}}{\partial^2 \sqrt{g_{xx}}\over\partial y^2}.
\end{equation}

Suppose that after embedding the sheet retains cylindrical symmetry.
Then along the outer rim of the flower, $\kappa$ is constant. One can
place a bound on $\kappa$. It 
cannot be greater in absolute value than $1/R$, where $R$ is the
radius of curvature at the rim, and it obtains this
value only when the edge of the cylinder has splayed out so that the
whole outer boundary shares a single tangent plane containing the
boundary. Thus the left hand side of Eq.~\ref{eq:G-B} obeys
\begin{equation}
-2\pi\leq\int ds\,\kappa=2\pi R\kappa\leq 2\pi
\label{eq:G-B2}
\end{equation}

Turning now to the right hand side of Eq.~\ref{eq:G-B}, one can
perform the integral in cylindrical coordinates. At any point indexed
by $y$, the radius of the cylinder is $R_\infty\sqrt{g_{xx}}$, where $R_\infty$
is the radius of curvature at the undistorted end of the flower. Along the $y$
direction, since $g_{yy}=1$, distances along the sheet are unchanged
by introduction of the metric, and one can write the integral over the
Gaussian curvature as
\begin{eqnarray}
\int dA\, K&=&\int_0^L dy\int d\theta R_\infty \sqrt{g_{xx}(y)}{1\over
 \sqrt{g_{xx}}}{\partial^2 \sqrt{g_{xx}}\over\partial y^2} 
\label{eq:G-B3}\\
&=&2\pi R_\infty {\partial \sqrt{g_{xx}}\over\partial y} \Big\vert_0^L
\label{eq:G-B4}\\
&=&-2\pi R_\infty {\partial \sqrt{g_{xx}(0)}\over\partial y} 
\label{eq:G-B5}
\end{eqnarray}

Returning to Eq.~\ref{eq:G-B}, one has the following condition:
\begin{equation}
\vert R_\infty {\partial \sqrt{g_{xx}(0)}\over\partial y} \vert\leq 1.
\label{eq:G-B6}
\end{equation}
For metrics with slopes less than this value, it is possible for a
cylindrical sheet to maintain cylindrical symmetry, but when the
gradient of the metric becomes too steep, a cylindrical sheet must
begin to buckle. 

Fig. \ref{fig:trumpet}(a) shows such a sheet at the point where the
metric has been chosen to make the inequality in Eq.~\ref{eq:G-B6} into an
equality, and the edges of the cylinder are splayed out as far as
they can go without buckling.  Fig. \ref{fig:trumpet}(b) provides an
example of a buckled cylinder, and the result suggests that the
number of ripples at the edge of a cylinder might be given roughly by
the quantity on the left hand side of Eq.~\ref{eq:G-B6}. I have not
pursued this idea further.

\section{Energy of a Thin Sheet}

I will now proceed to consider leaves, which is to say rectangular
strips of stretched material.  I will present some particular cases
where the shapes of rectangular strips with non--flat metrics can be
determined, even including very special cases where the shape is
described in terms of elementary functions. To begin, I will establish
in general the energy of a thin sheet with a metric that is not flat.

The energy of a thin sheet is
conventionally given by the F\"oppl--von 
K\'arm\'an equations, and is the sum of two terms, one involving
bending of the sheet and the other involving
stretching\cite{Landau.86,Mansfield.64}. This starting
point is inconvenient for three reasons. First, many
formulations employ coordinate systems that are not general enough to
encompass the folds and overhangs that occur in this problem. Second,
I will deal with sheets that have permanently been stretched, and the
equations must be generalized to encompass the deformation. Finally,
I need to move easily back and forth between numerical and analytical
approaches, and the numerical discretization of the conventional
equations is not simple.

I avoid all these problems by starting with a physical model of a
sheet based upon a discrete collection of interacting points. I
derive the continuum theory from the discrete model rather than by
discretizing continuum equations. Similar numerical techniques have been
employed frequently in studies of crumpled paper and tethered
membranes\cite{Lobkovsky.97,Seung.88}. 

Let $\vec u_i$ be a collection of mass points that interact with 
neighbors, and at rest form a thin flat sheet. Let $\vec\Delta_{ij}$
be equilibrium vector displacements between neighbors $i$ and
$j$. When the neighbors are not in equilibrium, the distance between
them is $u_{ij}=|\vec u_j-\vec u_i|$. Take the
energy corresponding to locations of the mass points to be
\begin{equation}
{\cal E}={{\cal K}\over 2 a} \sum_{\langle ij\rangle}
[u_{ij}^2-\Delta_{ij}^2]^2,\label{eq:energy}
\end{equation}
where ${\cal K}$ has dimensions of energy per volume, $a$ has
dimensions of length, and the sum is over pairs of neighbors. 

With Eq.~\ref{eq:energy} as a starting point, it is extremely easy to
see how to modify equations of elasticity to incorporate a new
metric. Write
\begin{equation}
{\cal E}={{\cal K}\over 2 a} \sum_{\langle ij\rangle}
[u_{ij}^2-\sum_{\alpha\beta}
\Delta^\alpha_{ij}g_{\alpha\beta}\Delta^\beta_{ij}]^2,\label{eq:energy2}  
\end{equation}
where $g_{\alpha\beta}$ is the metric tensor describing deformations
of the sheet. All that has happened, in short, is that the equilibrium
distance between material points has changed. My hypothesis is that
by specifying different metric tensors $g$, often ones with very
simple functional forms, one can describe all the buckling cascades we
have observed in the laboratory. In particular, since the plastic
sheets are torn uniformly in the $x$ direction, I will assume that
the metrics $g$ are constant in the $x$ direction, and vary only along
$y$. Often I will take $g$ to be diagonal, with $g_{xx}(y)$ a
smoothly varying function and $g_{yy}=1$.

\section{Continuum Limit}
The long--wavelength deformations of the points $\vec u_i$ correspond
to elastic deformations of a continuous sheet.  To form this correspondence, make
the replacement
\begin{equation}
\vec u_j\approx \vec u(\vec r_i)+ (\vec \Delta_{ij}\cdot\vec\nabla ) \vec
u(\vec r_i),
\end{equation}
where $\vec r_i$ is the location of $\vec u_i$ in an equilibrium of
Eq.~\ref{eq:energy}. Define the strain tensor 
\begin{equation}
\epsilon_{\alpha\beta}\equiv\mbox{$1\over 2$} \left [ \sum_\gamma
    {\partial u^\gamma\over \partial r_\alpha}
    {\partial u^\gamma\over \partial
    r_\beta}-g_{\alpha\beta}\right ].
\label{eq:strain}
\end{equation}
This definition reduces to the strain tensor of linear elasticity for
small deformations and flat metrics, and generalizes it appropriately
when deformations are large and the metric tensor $g$ differs from the
identity. In terms of the strain tensor, ${\cal E}$ can be rewritten as
\begin{equation}
{\cal E}={{\cal K}\over  a} \sum_i\sum_{j\mbox{\tiny\  nbr. of\ } i} \left [
  \sum_{\alpha\beta} \Delta_{ij}^\alpha \epsilon_{\alpha\beta}(\vec
  r_i)\Delta_{ij}^\beta \right ]^2.
\end{equation}

One has to specify a particular lattice in order to proceed
further. If one takes the points $\vec u_i$ to sit on two two--dimensional triangular
lattices, stacked over one another as in the first stage of forming an
hcp lattice of lattice constant $a$, the the energy takes the particular form
\begin{equation}
{\cal E}={{\cal K} a^3\over 24 } 
\sum_i \left [ 
\begin{array} {ll}
&19[\epsilon_{xx}+\epsilon_{yy}]^2+
    38[\epsilon_{xx}^2+\epsilon_{yy}^2+2\epsilon_{xy}^2]\\[3pt]
+&32\epsilon_{zz}^2+16[\epsilon_{yy}+\epsilon_{xx}]\epsilon_{zz}\\[3pt]
+&32[\epsilon_{yz}^2+\epsilon_{xz}^2]
    +8\sqrt{2}[\epsilon_{yz}(\epsilon_{yy}-\epsilon_{xx})+2\epsilon_{xy}\epsilon_{xz}] 
  \end{array}\right ]
\label{eq:continuum_energy}
\end{equation}

I will employ this continuum functional later to develop analytical
criteria for membrane shapes. Not all terms in it are equally important. However,
before arriving at specific deformations to insert into this functional, it
is difficult to tell which terms are large and which are small. I
therefore begin with a geometrical description of wrinkled sheets, and
then return to the question of which shapes minimize energy.

\section{Solvable Problem at Edge of Strip}

The full problem at hand is find the minimum energy state of
Eq.~\ref{eq:energy2} or Eq.~\ref{eq:continuum_energy} for a very long
strip of finite width and very small thickness $t$, subject to a
metric $g_{xx}(y)$ where $g_{xx}$ has some value $g_0$ at $y=0$ (left
hand side of lower panel in Figure \ref{fig:Membrane_setup}), and
decreases monotonically toward $1$ as $y$ approaches the other side
of the strip (right hand side of lower panel in Figure
\ref{fig:Membrane_setup}). It seems very unlikely that this problem
has in general an exact analytical solution. It is not even clear at
the outset whether minimum energy states inserted into the functional
Eq.~\ref{eq:continuum_energy} should produce the energies proportional to
$t$ that would be characteristic of stretching, or energies
proportional to $t^3$ that would be characteristic of bending.

Therefore, it is useful to find  cases where the problem can be solved
exactly. The motivation for the solvable problem comes by looking at
the lower panel in Figure \ref{fig:Membrane_setup}. Imagine making a
new infinitely long strip by slicing off the material a short distance
$w$ from the left hand side. If $w$ is small enough, then the metric
$g_{xx}$ within it should have the form of constant $g_0$ plus a term linear
in $y$. If the original length of this strip along the $x$ direction
was $L$, now its arc length is $\sqrt{g_0}L$. However, in order to be
able to join onto the rest of the strip on the right hand side, its
total extension in the $x$ direction must remain $L$. 

Therefore, I study a thin strip whose metric is
\begin{equation}
\sqrt{g_{xx}(y)}=\sqrt{g_0}(1-y/R);\quad g_{yy}=1
\label{eq:linear_metric}
\end{equation}
where $R$ is a constant. I look for solutions subject $\vec u(x,y,z)$
subject to the constraint that there be some period $\lambda$ for which
\begin{equation}
\vec u(x+\lambda,y,z)=\hat x \lambda+\vec u(x,y,z).
\label{eq:constraint}
\end{equation}
To obtain
the benefits of symmetry, also take 
\begin{equation}
y\in[-w/2,w/2]
\label{eq:center}
\end{equation}
so that the center line of the strip is at $y=0$, rather than one edge.

\begin{figure}
\begin{center}
\epsfysize4in\epsffile{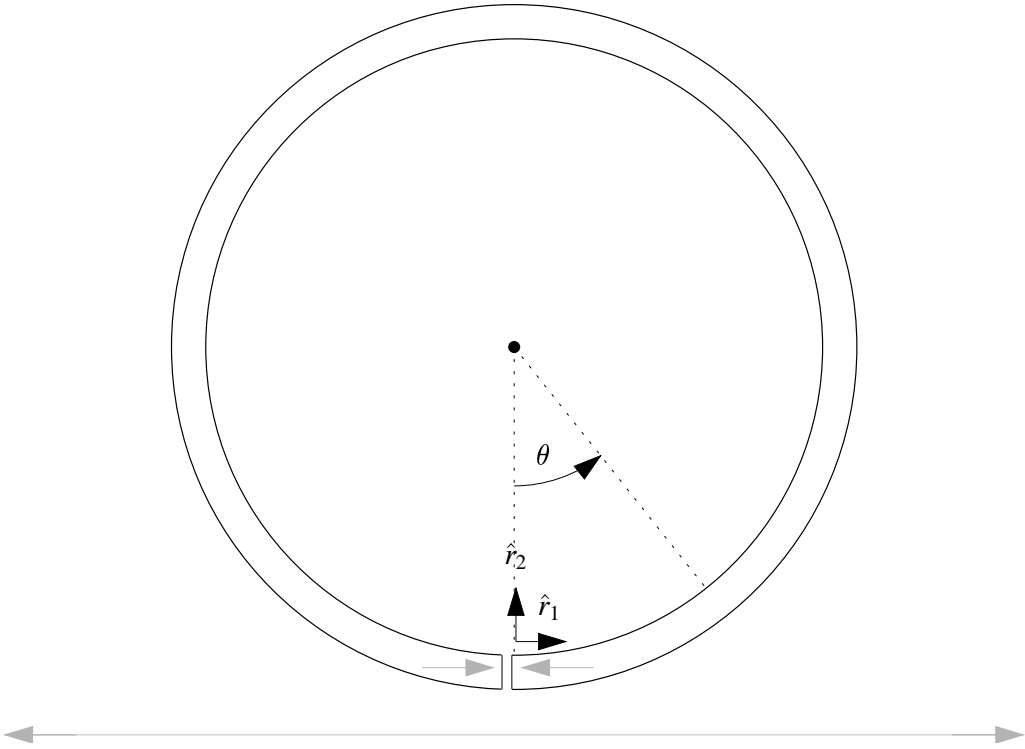}
\caption{\small Cut out the circular strip.
  Rotate the two ends so that the two gray arrows point 180$^\circ$
  away from one another, and place them over the lower line so that
  the arrows coincide.  The paper will assume a characteristic
  undulating shape that forms at the edges of leaves or stretched
  plastic, forming a solution close but not identical to that in Fig.
  \ref{fig:edge} with $\sqrt{g_0}=2$. }
\label{fig:circle}
\end{center}
\end{figure}

If a long strip of material is given metric Eq.~\ref{eq:linear_metric}
and no constraint is applied, then the minimum energy configuration is
easy to find. The material can relax completely by forming a ring of
radius $R$, which curls round and round in a circle. Thus one can
obtain some intuition about the problem by cutting out the paper
figure in Figure \ref{fig:circle} and pulling the ends apart
horizontally. Let $\hat r_1(\theta)$ and $\hat r_2(\theta)$ be two
unit vectors attached to the circular strip, where $\hat r_1$ points
around the circumference, and $\hat r_2$ points along the radius. Then
a class of low--energy deformations of the paper strip is produced by
introducing some bend at every angle $\theta$ that rotates only around
the current direction of $\hat r_2$.  Denote by $\omega(\theta)$ the
rate at which the strip rotates around $\hat r_2$ as a function of the
angle $\theta$. In a laboratory frame the directions of the unit
vectors are determined by
\begin{subequations}
\begin{eqnarray}
{\partial \hat r_1\over \partial \theta}&=&\hat r_2+\omega(\theta)\, \hat r_3 \label{eq:ode;a}\\
{\partial \hat r_2\over \partial \theta}&=&- \hat r_1
\label{eq:ode;b}\\[5pt]
\hat r_3&=&\hat r_1\times\hat r_2\label{eq:ode;c}
\end{eqnarray}
\label{eq:ode}
\end{subequations}
It is easy to check that $\hat r_1$, $\hat r_2$, and $\hat r_3$ remain
orthonormal unit vectors under the dynamics described by
Eqs.~\ref{eq:ode}. This simple observation leads to the first
important conclusion, which is that the constraint in
Eq.~\ref{eq:constraint} and the metric in Eq.~\ref{eq:linear_metric}
are incompatible if the strip undergoes bending alone. To show why,
relate coordinate $x$ and angle $\theta$ through
\begin{equation}
\sqrt{g_0}\, x=R\theta
\label{eq:angle_relation}
\end{equation}
and note that the location of the center line of the strip is given by 
\begin{equation}
\vec u(x,0,0)\equiv\vec l(\theta)\equiv R\int^\theta d\theta' \hat r_1(\theta').
\end{equation}
However, because of Eq.~\ref{eq:ode;b}, one can also write
\begin{equation}
\vec l(\theta)=- R \hat r_2(\theta).
\end{equation}
Since $\hat r_2$ is a unit vector, $\vec l$ cannot move more than
distance $2R$ from its starting point. Therefore it is impossible to
satisfy the constraint in Eq.~\ref{eq:constraint} when the only
bending permitted is around $\hat r_2$. 

\section{Solution with Constant Bending}
Despite the fact that bending about $\hat r_2$ alone cannot solve the
physical problem at hand,
there is a solution of Eqs.~\ref{eq:ode} needed for further mathematical
development. This solution is found when $\omega$ is
constant. The equations are not obviously solvable even in this simple
case, since they are
nonlinear, but the solution can be expressed in terms of
elementary functions. Let
\begin{subequations}
\begin{eqnarray}
\hat r_1&=&-\mbox{Re}[{i\vec Be^{i\theta/\theta_0}\over \theta_0}]\\
\hat r_2&=&\vec A+\mbox{Re}[\vec Be^{i\theta/\theta_0}],
\end{eqnarray}
\label{eq:solform}
\end{subequations}
where $\vec A$ is real but $\vec B$ is complex. With this choice of
$\hat r_1$, one immediately satisfies \ref{eq:ode;b}. In order for
$\hat r_1$ to be a unit vector,
\begin{equation}
\vec B\cdot\vec B=0\quad\mbox{and}\quad\vec B\cdot\vec
B^*={2\theta_0^2}.
\label{eq:Bcond}
\end{equation}
Now substitute Eqs.~\ref{eq:solform} into Eqs.~\ref{eq:ode;c} and
\ref{eq:ode;a}. They are satisfied if
\begin{equation}
\mbox{Re}[{B\over\theta_0^2}e^{i\theta/\theta_0}]={-i\omega \over
  4\theta_0}\left [
e^{i\theta/\theta_0}\vec B\times A-\mbox{c.c.}+2\vec B\times\vec
  B^*\right ] +\vec A+\mbox{Re}[Be^{i\theta/\theta_0}].
\end{equation}
Matching up coefficients of different powers of $\exp[i\theta/\theta_0]$
gives
\begin{subequations}
\begin{eqnarray}
\vec A&=&{i\omega\over 2\theta_0}\vec B\times\vec B^* \label{eq:Bcond2;a}
\\
\vec B(\theta_0^{-2}-1)&=& {-i\omega\over 2\theta_0}\vec B\times \vec A.
\label{eq:Bcond2;b}
\end{eqnarray}
\label{eq:Bcond2}
\end{subequations}
Substituting Eq.~\ref{eq:Bcond2;a} into  Eq.~\ref{eq:Bcond2;b} and
using Eq.~\ref{eq:Bcond} gives
\begin{equation}
\theta_0^2={1\over \omega^2+1}.
\end{equation}
Choosing any $\vec B$ that satisfies Eq.~\ref{eq:Bcond}, one therefore
has a solution, where $\vec A$ is determined by
Eq.~\ref{eq:Bcond2;a}. In particular, when $\hat r_1(0)=\hat x$ and
$\hat r_2(0)=\hat y$,
one has explicitly
\begin{subequations}
\begin{eqnarray}
\hat r_1&=&\cos(\theta/\theta_0)\hat x+\theta_0\sin(\theta/\theta_0)\hat
y +\theta_0 \omega\sin(\theta/\theta_0)\hat z \\
\hat r_2&=&-\theta_0\sin(\theta/\theta_0)\hat
x+(1-\theta_0^2(1-\cos(\theta/\theta_0))) \hat
y-\omega\theta_0^2(1-\cos(\theta/\theta_0)) \hat z \\
\hat r_3&=&-\theta_0\omega\sin(\theta/\theta_0)\hat
x-\theta_0^2\omega[1-\cos(\theta/\theta_0)]\hat
y+(\theta_0^2+[1-\theta_0^2]\cos(\theta/\theta_0))\hat z.
\label{eq:so;c}
\end{eqnarray}
\label{eq:sol}
\end{subequations}
and
\begin{subequations}
\begin{eqnarray}
l^x&=&{R\sin( \theta/\theta_0)\theta_0}\\
l^y&=&{R(1-\cos(\theta/\theta_0))\theta_0^2}\\
l^z&=&{R\omega(1-\cos(\theta/\theta_0))\theta_0^2}
\end{eqnarray}
\end{subequations}

\section{Solution with Constant Bending and Torsion}
Since bending around $\hat r_2$ is not a general enough deformation of
the strip to satisfy constraint Eq.~\ref{eq:constraint}, one must proceed
next to consider torsion; that is, bends around $\hat r_1$.
Let $\dot\phi$ describe the rate at which twisting around $\hat r_2$
occurs, and let $\dot \psi$ describe the rate at which twisting around
$\hat r_1$ occurs. A first case allowing exact solution is when 
$\dot \phi$ and $\dot\psi$ are constant. Then Eqs.~\ref{eq:ode}
become
\begin{subequations}
\begin{eqnarray}
{\partial \hat r_1\over \partial \theta}&=&\dot\phi \hat r_1\times\hat
r_2+\hat r_2 \label{eq:ode2;a}\\
{\partial \hat r_2\over \partial \theta}&=&\dot\psi \hat r_1\times\hat
r_2- \hat r_1.
\label{eq:ode2;b}
\end{eqnarray}
\label{eq:ode2}
\end{subequations}
Define
\begin{eqnarray}
\begin{pmatrix}
\hat s_1\\
\hat s_2
\end{pmatrix}
={1\over \omega}  
\begin{pmatrix} 
\dot\phi &\dot\psi \\
 -\dot \psi &\dot \phi 
\end{pmatrix}
\begin{pmatrix} \hat r_1 \\
 \hat r_2
\label{eq:s}
\end{pmatrix}
 \\
\noalign{\vbox{where}}
\omega=\sqrt{\dot\phi^2+\dot\psi^2}
\end{eqnarray}
Rewriting Eqs.~\ref{eq:ode2} in terms of these new variables, they
become 

\begin{subequations}
\begin{eqnarray}
{\partial \hat s_1\over \partial \theta}&=&\hat s_2+\omega \hat s_1\times\hat
s_2 \label{eq:ode3;a}\\
{\partial \hat s_2\over \partial \theta}&=&- \hat s_1.
\label{eq:ode3;b}
\end{eqnarray}
\label{eq:ode3}
\end{subequations}

Eqs. \ref{eq:ode} and Eqs. \ref{eq:ode3} are identical, except that
the latter involve $s$ instead of $r$. Therefore, choosing a coordinate system where
$\hat x'$ points along $\hat s_1(0)$ and $\hat y'$ points along
$s_2(0)$, the $\hat x'$, $\hat y'$ and $\hat z'$ coordinates of $\hat
s_1$ and $\hat s_2$ are given once again by  Eqs.~\ref{eq:sol}.

\section{Solution with Oscillating Bending and Torsion}
This solution still does not solve the physical problem at hand,
because the strip twists endlessly around the original $\hat x$
axis. In order to avoid twisting the strip, the torsion $\dot \psi$ must
oscillate between positive and negative values. A solution of this
type can be found by taking 
\begin{subequations}
\begin{eqnarray}
\dot \phi&=&\omega\cos\alpha\theta\\
\dot\psi&=&\omega\sin\alpha\theta
\label{eq:alpha}
\end{eqnarray}
\end{subequations}

Now the equations describing rotations of the strip unit vectors are
nonlinear and have non--constant coefficients. However, they can still
be solved. Once again, define a new coordinate system with
Eq.~\ref{eq:s}.  Substituting  Eqs.~\ref{eq:alpha} into
Eqs.~\ref{eq:ode} gives now
\begin{subequations}
\begin{eqnarray}
{\partial \hat s_1\over \partial \theta}&=&\hat s_2(1+\alpha)+\omega \hat s_1\times\hat
s_2 \label{eq:ode4;a}\\
{\partial \hat s_2\over \partial \theta}&=&- \hat s_1(1+\alpha).
\label{eq:ode4;b}
\end{eqnarray}
\label{eq:ode4}
\end{subequations}
Defining 
\begin{equation}
\theta'=(1+\alpha)\theta,\quad
\omega'=\omega/(1+\alpha)\quad\mbox{and}\quad {\theta'_0}^2=
{1\over{1+{\omega'}^2}} 
\label{eq:pdefs}
\end{equation}
one again recovers Eqs.~\ref{eq:ode}, and its solution
Eqs.~\ref{eq:sol} in terms of primed variables. That is,
\begin{subequations}
\begin{eqnarray}
\hat s_1&=&\cos(\theta'/\theta_0')\hat x+\theta_0'\sin(\theta'/\theta_0')\hat
y +\theta'_0 \omega'\sin(\theta'/\theta'_0)\hat z \\
\hat s_2&=&-\theta'_0\sin({\theta'\over\theta'_0})\hat
x+(1-{\theta_0'}^2(1-\cos({\theta'\over\theta_0'}))) \hat
y-\omega'{\theta_0'}^2(1-\cos({\theta'\over\theta_0'})) \hat z \\
\hat s_3&=&-\theta_0'\omega'\sin({\theta'\over \theta_0'})\hat
x-{\theta_0'}^2\omega'[1-\cos({\theta'\over \theta_0'})]\hat
y+({\theta_0'}^2+[1-{\theta_0'}^2]\cos({\theta'\over \theta_0'}))\hat z.
\label{eq:solb;c}
\end{eqnarray}
\label{eq:solb}
\end{subequations}

Finally the solutions can correspond to the shapes at the edge of
wrinkled sheets. To see when they are physically acceptable, one has
to invert all the linear transforms and write the expression for $\hat
r_1$ explicitly. It is
\begin{subequations}
\begin{eqnarray}
\hat r_1^x&=&
\cos\alpha\theta\cos(\theta'/\theta'_0)+\theta'_0\sin\alpha\theta\sin(\theta'/\theta'_0)
\\
\hat r_1^y&=&
\theta'_0\cos\alpha\theta
\sin(\theta'/\theta'_0)-\sin\alpha\theta(1-{\theta'_0}^2(1-\cos\theta
'/\theta'_0)) \\
\hat r_1^z&=&
\theta'_0\omega'\cos\alpha\theta\sin(\theta'/\theta'_0)+\omega'
{\theta'_0}^2\sin\alpha\theta (1-\cos(\theta'/\theta'_0))
\end{eqnarray}
\label{eq:r}
\end{subequations}
To satisfy the constraint Eq.~\ref{eq:constraint}, when one integrates $\hat r_1$ to get
$\vec l$, the $x$ component must increase indefinitely, while the $y$
and $z$ components must oscillate. The solution acts this way if and
only if 
\begin{equation}
\alpha\theta=\pm\theta'/\theta'_0\label{eq:a_condition}
\end{equation}
The two signs in Eq.~\ref{eq:a_condition} produce identical solutions,
as Eq.~\ref{eq:r} is invariant under
$\theta'_0\rightarrow-\theta'_0$. Adopting the minus sign, one has
\begin{eqnarray}
\alpha&=&-{(1+\omega^2)/2}\label{eq:a_value}\\
\Rightarrow \theta'_0&=& {1-\omega^2\over 1+\omega^2}
\end{eqnarray}
The vectors $\hat r_1\dots\hat r_3$ are
\begin{subequations}
\begin{eqnarray}
\hat  r_1^x &=& \cos^2(\alpha\theta)-\sin^2(\alpha\theta) \theta_0'\\
\hat  r_1^y &=& \big[1-\cos(\alpha\theta)\big] \sin(\alpha\theta) \theta_0'^2-\cos(\alpha\theta) \sin(\alpha\theta) \theta_0'-\sin(\alpha\theta)\\
\hat  r_1^z &=& \Big(\big[1-\cos(\alpha\theta)\big] \sin(\alpha\theta) \theta_0'^2-\cos(\alpha\theta) \sin(\alpha\theta) \theta_0'\Big) \omega'\\[4pt]
\hat  r_2^x &=& \cos(\alpha\theta) \sin(\alpha\theta) \theta_0'+\cos(\alpha\theta) \sin(\alpha\theta)\\
\hat  r_2^y &=& \big[\cos^2(\alpha\theta)-\cos(\alpha\theta)\big] \theta_0'^2-\sin^2(\alpha\theta) \theta_0'+\cos(\alpha\theta)\\
\hat r_2^z &=& \Big(\big[\cos^2(\alpha\theta)-\cos(\alpha\theta)\big] \theta_0'^2-\sin^2(\alpha\theta) \theta_0'\Big) \omega'\\[4pt]
\hat r_3^x &=& \sin(\alpha\theta) \theta_0' \omega'\\
\hat r_3^y &=& (\cos(\alpha\theta)-1) \theta_0'^2 \omega'\\
\hat  r_3^z &=& -\cos(\alpha\theta)
\theta_0'^2+\cos(\alpha\theta)+\theta_0'^2
\end{eqnarray}
\label{eq:vectors}
\end{subequations}
        
The wavelength of the pattern is given by the angle through which
$\theta$ travels so that the smallest nonzero Fourier component of the pattern
goes through one period, meaning that $\theta$ travels through $2\pi/\alpha$.
One now finds that the wavelength of the pattern $\lambda=l(2\pi/\alpha)$ is
\begin{equation}
\lambda={2\pi\over|\alpha|} R(1-\theta'_0)/2={4\pi R\omega^2\over (1+\omega^2)^2}.
\end{equation}
Since the strip travels an arclength $2\pi R/\alpha$ when $\theta$ goes through
$2 \pi/\alpha$, one can find $L/\lambda\equiv\sqrt{g_0}$:
\begin{equation}
\sqrt{g_{0}}=1+1/\omega^2 \Rightarrow
\omega={1\over\sqrt{\sqrt{g_0}-1}}.
\label{eq:metric_solution}
\end{equation}

Furthermore, using Eq.~\ref{eq:linear_metric}, one can determine the
wavelength $\lambda$ from
\begin{equation}
{2\pi \over \lambda}={{g_{0}}\over 2R(\sqrt{g_{0}}-1)}
={1\over 4(\sqrt{g_{0}}-1)} \left
  |{\partial {g_{xx}}
  \over \partial y}\right |_{y=0}.
\label{eq:wavelength}
\end{equation}
Since $\omega$ and $R$, or equivalently $\lambda$, are the only
parameters describing the solution, it has been determined completely
by $\sqrt{g_0}$ and the slope of the metric.  In particular, the
wavelength of the pattern has been determined. 

The functional forms of bending and torsion angles in
Eq.~\ref{eq:alpha} were simply pulled out of a hat. One must
wonder whether in fact the resulting shapes minimize the energy
functional in Eq.~\ref{eq:energy2} with which the problem
began. This question will be studied in later sections, and the
answer is ``sometimes.'' That is, there can exist energy functionals
and boundary conditions for which these solutions are energy
minimizers. However, for the particular functional in
Eq. \ref{eq:continuum_energy}, the solutions found in this section are
excellent approximations but never exact energy minimizers. Furthermore,
even for those functionals whose energy they can minimize, they do so
only for a restricted subset of boundary conditions one can naturally
apply to a strip. Returning to the paper strip in
Fig.~\ref{fig:circle}, the solution described by Eq.~\ref{eq:vectors}
is only legitimate when the two ends of the strip are placed at a
particular distance as specified in the caption. However, if one
holds the strip in ones hands, it is easy to slide its ends
horizontally back and forth. The energy--minimizing shapes resulting
from this process certainly exist, but happen not to be in the
analytical family following from Eq.~\ref{eq:alpha}. Finally, one can
ask what happens if one takes a strip of length $L\gg R$ and allows it
to curl up into an energy--minimizing shape. Will it ever adopt a
periodic structure with the period given exactly by
Eq.~\ref{eq:wavelength}? The answer to this question is probably
``No,'' but it has not yet fully been settled

\begin{figure}[H]
\centerline{\epsfxsize=4.5in\epsffile{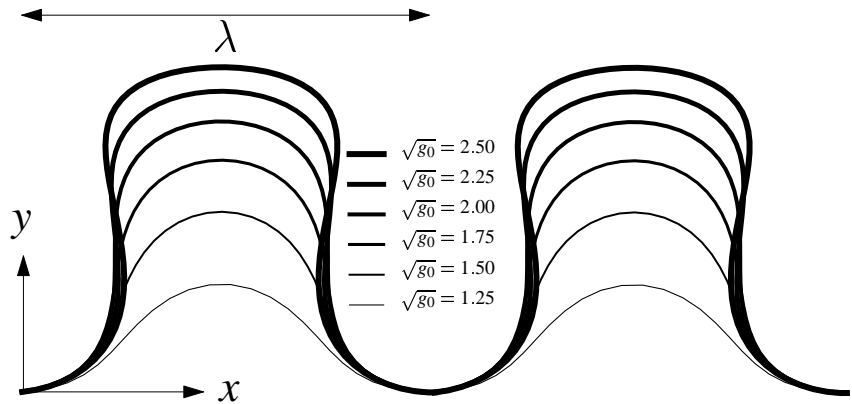}}
\caption{\small Height of edge $l^y(\theta)$ versus length of edge
  $l^x(\theta)$ for solution given in Eq.~\ref{eq:r} and various
  values of $\sqrt{g_0}$. }
\label{fig:edge}
\end{figure}

Some images of solutions from Eq. \ref{eq:vectors} appear in Fig.~\ref{fig:edge}. The radius $R$
is adjusted in each case so that the wavelength $\lambda$ remains
constant. Drawing a number of such pictures, one finds that there is
an upper limit of $\sqrt{g_0}\approx 5.61$ for solutions of this
type. When $\sqrt{g_0}$ is larger than this value, the solution
collides with itself in the center of the wave and becomes
self--intersecting. 

\section{Implications for Energy Functional}

Assuming that displacements are of the form given in Eq.~\ref{eq:r},
it is possible to return to the functional
Eq.~\ref{eq:continuum_energy} and decide which terms dominate the
energy of a stretched sheet. From the customary theory of thin sheets
one expects the energy to come from the sum of two terms. The first term
is the stretching energy of the sheet, and is of order $t$, where $t$
is the thickness of the sheet; in Eq.~\ref{eq:continuum_energy}, the
stretching energy would arise from the terms on the top of the right
hand side that involve no derivatives with respect to $z$. The second
term is the bending energy of the sheet, and it is proportional to
$t^3$.

This expectation is correct. Bending and stretching energies dominate
the energy functional. Which of them dominates is a subtle matter. The
stretching energy is proportional to the thickness of the sheet $t$
while the bending energy is proportional to the thickness of the sheet
cubed, $t^3$. In the limit of thin sheets, one would think that the
stretching energy would be dominant. However, matters are not that
simple, because the energy is proportional to the width of the strip
$w$ to the fifth power, $w^5$. The limit I believe is most
interesting in practice is
\begin{equation}
\left ({w\over R}\right )^2\ll {t\over R}\ll {w\over R}.
\label{eq:inequalities}
\end{equation}
In this limit the bending energy proportional to $t^2/R^2$ dominates
the energy, while the stretching energy, proportional to $w^4/R^4$,
can be neglected.

To obtain the energy formally, take $\vec u$ in the form
\begin{eqnarray}
\vec u(x,y,z)&=&\vec l(\theta)+\hat r_2 y+\hat r_3 z\\
&+& \left [q^{\scriptscriptstyle (1)}_{zz}  z^2
+q^{\scriptscriptstyle (1)}_{zy}  z y
+q^{\scriptscriptstyle (1)}_{yy}  y^2 \right] \hat r_1/R\\
&+&\left [
q^{\scriptscriptstyle (2)}_{zz}  z^2
+q^{\scriptscriptstyle (2)}_{zy} z y
+q^{\scriptscriptstyle (2)}_{yy} y^2 \right ] \hat r_2/R\\
&+&
\left[ q^{\scriptscriptstyle (3)}_{zz}  z^2
+q^{\scriptscriptstyle (3)}_{zy}  z y
+q^{\scriptscriptstyle (3)}_{yy}  y^2\right ] \hat r_3/R
\label{eq:uform2}
\end{eqnarray}
with $\theta$ related to $x$ by Eq.~\ref{eq:angle_relation}.

If all the coefficients $q$ were zero, one would have
\begin{subequations}
\begin{eqnarray}
{\partial \vec u\over \partial x}&=&{\sqrt{g_0}\over R}{\partial\over
  \partial\theta} \left [
\vec l(\theta)+\hat r_2 y+\hat r_3 z
\right ]\\
&=&\sqrt{g_0}[ \hat r_1+{y\over R}{\partial \hat r_2\over \partial
  \theta}
  +{z\over R}{\partial \hat r_3\over \partial \theta} ]\\
&=&\sqrt{g_0}[\hat r_1(1-{y\over R}-\dot\phi {z\over R})  -{z\over R}
  \dot\psi \hat
  r_2+{ y\over R}\dot\psi \hat r_3] \label{eq:uderivs;a}
\\
{\partial \vec u\over \partial y}&=& \hat r_2  \label{eq:uderivs;b}
\\
{\partial\vec u\over \partial z}&=& \hat r_3.
\label{eq:uderivs;c}
\end{eqnarray}
\label{eq:uderivs}
\end{subequations}
However, the functions $q$ are needed to allow small contractions of the
membrane so as to minimize its energy. The minimizations are performed
in the following way: first find the term in
Eq.~\ref{eq:continuum_energy} proportional to $y^2/R^2$. This term can be
made to vanish through a unique choice of $q^{\scriptscriptstyle (1)}_{zy}$,
$q^{\scriptscriptstyle (2)}_{zy}$,
$q^{\scriptscriptstyle (3)}_{zy}$, $q^{\scriptscriptstyle (1)}_{yy}$ ,and $q^{\scriptscriptstyle (2)}_{yy}$,
where $q^{\scriptscriptstyle (1)}_{zy}=-\dot\psi$,
$q^{\scriptscriptstyle (2)}_{zy}=-2q^{\scriptscriptstyle (3)}_{yy}$, and the others vanish. The next term to
consider is the term proportional to $z^2/R^2$, as this is larger than
the term proportional to $y^4/R^4$. Minimizing the term proportional
to $z^2$ determines the following values for the variables $q$:

\begin{subequations}
\begin{eqnarray}
q^{\scriptscriptstyle (1)}_{zz}&=&\sqrt{2}\dot\psi/4,\quad
q^{\scriptscriptstyle (1)}_{zy}=-\dot\psi,\quad
q^{\scriptscriptstyle (1)}_{yy}=0\\[5pt]
q^{\scriptscriptstyle (2)}_{zz}&=&\sqrt{2}\dot\phi/6,\quad
q^{\scriptscriptstyle (2)}_{zy}=\dot\phi/3,\quad
q^{\scriptscriptstyle (2)}_{yy}=0\\[5pt]
q^{\scriptscriptstyle (3)}_{zz}&=&\dot\phi/12,\quad\quad
q^{\scriptscriptstyle (3)}_{zy}=0,\quad\quad
q^{\scriptscriptstyle (3)}_{yy}=-\dot\phi/6.
\label{eq:qvals}
\end{eqnarray}
\end{subequations}

Inserting the expressions for $\epsilon_{\alpha\beta}$ into
Eq.~\ref{eq:energy2}, one has the following results. Since there are
are only two layers in the $z$ direction, carry out that sum
explicitly, but convert sums in the $x$ and $y$ direction to
integrals, using the fact that the area per particle is
$\sqrt{3/4}a^2$. The leading term proportional to the sheet thickness
is of the form
\begin{equation}
{\cal K} {a L w^5\over R^4}
{1\over L} \int_0^L dx\quad[\mbox{many terms!}]
\end{equation}
but the term that dominates the energy when inequalities
Eq.~\ref{eq:inequalities} hold is
\begin{equation}
{\cal K} {a t^2 L w\over R^2}
{1\over L} \int_0^L dx
{{\left(3 g_0 {\dot\psi}^{2}+2 g_0^2{\dot\phi}^{2}\right)
 }\over\sqrt{3 }}.
\label{eq:dominanta}
\end{equation}
When Eq.~\ref{eq:dominanta} dominates the energy, the strip is wide
enough so that $w\gg t$, but not so wide that it is favorable to begin
buckling or forming fine structure in the $y$ direction.

The various powers of $g_0$ appearing in Eq.~\ref{eq:dominanta} 
arise because the microscopic lattice
underlying the calculation has been stretched, and its elastic
properties made anisotropic by the new metric. The calculations are
producing a specific description for how stretching material in the
$x$ direction by a factor of $\sqrt{g_0}$ changes its elastic
properties. This description is unlikely to apply to the polymeric
materials in which most experiments are conducted. For this reason,
direct numerical studies of the bending of strips have been carried
out in such a way that $g_0$ in  Eq.~\ref{eq:dominanta} should be set
to 1. Rather than starting with a strip of length $L=\lambda$, stretching it
to length $\lambda\sqrt{g_0}$, and then introducing a gradient in the metric
along $y$, the numerical studies start with a strip of length $L$,
introduce a linear gradient of slope $1/R$ in the metric along $y$, and then
constrain the ends of the strip to sit at distance
$\lambda=L/\sqrt{g_0}$. That is, the constraint at the ends of the
strip instead of Eq.~\ref{eq:constraint} is
\begin{equation}
\vec u(x+L,y,z)=\hat x \lambda+\vec u(x,y,z).
\label{eq:constraint2}
\end{equation}
The energy of a strip prepared in this way is
\begin{equation}
{\cal K} {a t^2 L w\over R^2}
{1\over L} \int_0^L dx
{{\left(3  {\dot\psi}^{2}+2 {\dot\phi}^{2}\right)
 }\over\sqrt{3 }}.
\label{eq:dominant}
\end{equation}

\section{Euler--Lagrange Equations for Minimum of Elastic Energy}

The solution of Eq.~\ref{eq:r} was obtained by assuming that
$\dot\phi$ and $\dot\psi$ have the forms given in Eq.~\ref{eq:alpha}.
These choices for the torsion and bending of the strip did make it
possible to find a shape for the strip that was consistent with the
constraint in Eq.~\ref{eq:constraint}, and with a few other physical
considerations.

Varying $\dot \phi$ and $\dot \psi$ away
from the arbitrary forms obtained by guessing leads to additional
solutions, generally of lower energy. Depending upon the
precise form of the continuum energy functional, and the boundary
conditions to which the problem is subjected, the solutions in Eq.~\ref{eq:vectors}
can turn out to be exact, and in many other cases are excellent
approximations to exact solutions.

To look for energy--minimizing shapes, I will use
Eq.~\ref{eq:dominant} to express the energy of a strip, and will
rewrite it as
\begin{equation}
{\cal E}=\int d\theta\,{C_1\over2}\dot\phi^2+{C_2\over 2}\dot\psi^2.
\label{eq:MIN1}
\end{equation}
If $C_1$ is equal to $C_2$ then the analytic solution  of
Eq.~\ref{eq:vectors} can minimize
the energy for certain boundary conditions. According to
Eq. \ref{eq:dominant}, in fact the constants are
in the ratio of 2 to 3, and since they are not equal, the solution
is approximate, although for some boundary conditions the
approximation is excellent.

In explaining the calculations leading to these conclusions, it is
useful to derive the equations of motion Eq.~\ref{eq:ode} in a
different way. Let $\hat r_1(\theta)\dots\hat r_3(\theta)$ be
orthonormal vectors that evolve as a function of $\theta$. Letting
primes denote derivatives with respect to $\theta$, one has in general
that
\begin{equation}
\hat r'_i=\sum_j (\hat r'_i\cdot\hat r_j) \hat r_j
\label{eq:MIN2}
\end{equation}
This equation of motion is subject to a number of constraints. First,
each vector must retain unit magnitude. Therefore
\begin{subequations}
\begin{equation}
|\hat r_i|^2=0\Rightarrow \hat r'_i\cdot\hat r_i=0.
\label{CONSTRAINT1}
\end{equation}
Second, since the vectors are orthogonal, for $i\neq j$
\begin{equation}
\hat r_i\cdot\hat r_j=0.
\label{CONSTRAINT2}
\end{equation}
Third, the condition that $\hat r_1$ and $\hat r_2$ curl around one
another because of the gradient of metric, but otherwise be rigid
against twisting around $\hat r_3$ means that
\begin{equation}
\hat r_1\cdot\hat r_2'=-1\quad\mbox{or}\quad\hat r_1'\cdot\hat r_2=1.
\label{CONSTRAINT3}
\end{equation}
\label{CONSTRAINTS}
\end{subequations}

There are no constraints upon $\hat r_3\cdot\hat r_1'$ or $\hat
r_3\cdot\hat r_2'$; define these functions to be
\begin{subequations}
\begin{eqnarray}
\dot\phi&\equiv&\hat r_3\cdot\hat r_1'\\
\dot\psi&\equiv&\hat r_3\cdot\hat r_2'.
\end{eqnarray}
\label{PHIDEF}
\end{subequations}
Then Eqs.~\ref{eq:ode} follow from Eq.~\ref{eq:MIN2} just by employing
Eqs.~\ref{CONSTRAINTS} and \ref{PHIDEF}.

Once vectors $\hat r_1\dots\hat r_3$ are subject to the constraints in
Eqs.~\ref{CONSTRAINTS}, the energy Eq.~\ref{eq:MIN1} can be rewritten,
apart from an overall additive constant, as 
\begin{equation}
{\cal E}=\int d\theta\, {C_1\over 2} |\hat r_1'|^2+{C_2\over 2}|\hat
r_2'|^2 
\end{equation}
Imposing the final constraint
\begin{equation}
\int_0^{\theta_f} d\theta \hat r_1=\lambda\hat x,
\end{equation}
one needs to extremize the functional
\begin{equation}
\int d\theta\, {C_1\over 2} |\hat r_1'|^2+{C_2\over 2}|\hat
r_2'|^2 +\vec p\cdot\hat r_1(\theta)-\mbox{$1\over 2$}u_1(\theta)|\hat
r_1^2|-
u_2(\theta)\hat r_1\cdot \hat r_2-
\mbox{$1\over 2$}u_1(\theta)|\hat r_2^2|
-u_4(\theta)\hat r_1\cdot\hat r_2'.
\label{eq:FUNCTIONAL}
\end{equation}
The constant vector $\vec p$ and the functions $u_1(\theta)\dots
u_4(\theta)$ are Lagrange multipliers enforcing the constraints.

There may be some concern over the legitimacy of using equations of
constraint in order to simplify the form of the energy functional
before carrying out the variation. The results will not be reported
here in detail, but the entire variational procedure has been repeated
using $\vec r_3\cdot\vec r_1'$ and $\vec r_3\cdot\vec r_2'$ as the
definitions of $\dot\phi$ and $\dot\psi$, and enforcing directly the
constraints $\vec r_1'=\dot\phi\vec r_1\times\vec r_2+\vec r_2$ and
$\vec r_2'=\dot\psi\vec r_1\times\vec r_2-\vec r_1$. The algebra is
more involved than what is reported below, but the results are
identical.

Taking the variation of the functional Eq. \ref{eq:FUNCTIONAL} with respect to $\vec r_1$ and $\vec r_2$ gives
\begin{subequations}
\begin{eqnarray}
-C_1\vec r_1''+\vec p -u_1\vec r_1-u_2\vec r_2-u_4\vec r_2'=0\\
-C_2\vec r_2''-u_2\vec r_1-u_3\vec r_2+u_4 \vec r_1'+u_4'\vec r_1=0.
\label{eq:VAR1}
\end{eqnarray}
\end{subequations}
Now employing Eq.~\ref{eq:ode}, one can rewrite Eqs.~\ref{eq:VAR1} as 
\begin{subequations}
\begin{eqnarray}
C_1[-(\dot\phi^2+1)\vec r_1-\dot\phi\dot\psi\vec
r_2+[\dot\phi'+\dot\psi]]+u_1\vec r_1+u_2\vec r_2+u_4[\dot\psi\vec
r_3-\vec r_1]=\vec p \label{eq:VAR2;a}\\
C_2[-\dot\phi\dot\psi\vec r_1-(\dot\psi^2+1)\vec
r_2+(\dot\psi'-\dot\phi)\vec r_3]+u_2\vec r_1+u_3\vec
r_2-u_4[\dot\phi\vec r_3+\vec r_2]-u_4'\vec r_1=0
\label{eq:VAR2;b}
\end{eqnarray}
\label{eq:VAR2}
\end{subequations}
One obtains six scalar equations by taking the dot product of
Eqs.~\ref{eq:VAR2} with $\hat r_1$, $\hat r_2$ and $\hat r_3$ in turn,
using the fact that they are orthonormal once all the Lagrange
multipliers are chosen properly. The only time that $u_1$ appears is
when one takes the dot product of Eq.~\ref{eq:VAR2;a} with $\hat r_1$,
and similarly the only time that $u_3$ appears is when one takes the
dot product of Eq.~\ref{eq:VAR2;b} with $\hat r_2$. Therefore, $u_1$
and $u_3$ can be taken to be whatever is needed to satisfy these two
equations, which need not even be written down. The remaining four
equations are
\begin{subequations}
\begin{eqnarray}
u_2&=&C_2\dot\phi\dot\psi+u_4'\\
u_4\dot\phi&=&C_2[\dot\psi'-\dot\phi]\\
u_2&=&\vec p\cdot\hat r_2+C_1\dot\phi\dot\psi\\
u_4\dot\psi&=&\vec p\cdot\hat r_3-C_1[\dot\phi'+\dot\psi]
\end{eqnarray}
\label{eq:VAR3}
\end{subequations}
Eliminating $u_2$ allows one finally to write a complete set of Euler--Lagrange
equations determinining minima of the energy:
\begin{subequations}
\begin{eqnarray}
u_4'&=&\vec p\cdot\hat r_2+[C_1-C_2]\dot\phi\dot\psi\label{eq:VAR3;a}
\\
\dot\psi'&=&u_4\dot\phi/C_2+\dot\phi\label{eq:VAR3;b}
\\
\dot\phi'&=&\big [\vec p\cdot(\hat r_1\times \hat r_2)-u_4\dot\psi\big]/C_1-\dot\psi\label{eq:VAR3;c}
\\
\hat r_1'&=&\dot\phi (\hat r_1\times \hat r_2)+\hat r_2\label{eq:VAR3;d}
\\
\hat r_2'&=&\dot\psi (\hat r_1\times \hat r_2)-\hat r_1.\label{eq:VAR3;e}
\end{eqnarray}
\label{eq:VAR4}
\end{subequations}
Notice that all the terms in Eq.~\ref{eq:vectors} have definite parity
under $\theta\rightarrow-\theta$. Assuming that all energy minima have
the same symmetry, $\vec p$ must point along $x$, $\dot\phi$ and $u_4$
must be even, and $\dot \psi$ must be odd. The solutions of
Eq.~\ref{eq:VAR4} are indexed by four constants, the initial values
of $\dot\phi$ and $u_4$, and by $p_x$, and finally the angle
$\theta_f$ to which the solutions are to be integrated. 

To this point, the manipulations make use of Eq.~\ref{eq:ode}, but do
not use any features of the particular forms of $\dot\phi$ and
$\dot\psi$ chosen in Eq.~\ref{eq:alpha}. Do $\dot\phi$ and $\dot\psi$
chosen according to Eq. \ref{eq:alpha} solve
Eqs.~\ref{eq:VAR4}? Inserting these forms of $\dot\phi$ and
$\dot\psi$, one quickly finds that  
\begin{eqnarray}
u_4&=&C_2(\alpha-1), \quad\mbox{and}\quad u_4'=0\\
&&\Rightarrow (C_2-C_1)\omega^2\sin\alpha\theta\cos\alpha\theta=\vec
p\cdot\hat r_2\\
&&(C_1-C_2)(1-\alpha)\omega\sin\alpha\theta=\vec p\cdot\hat r_3.
\end{eqnarray}

\begin{figure}[!ht]
\begin{center}
(a)\epsfxsize=4in\epsffile{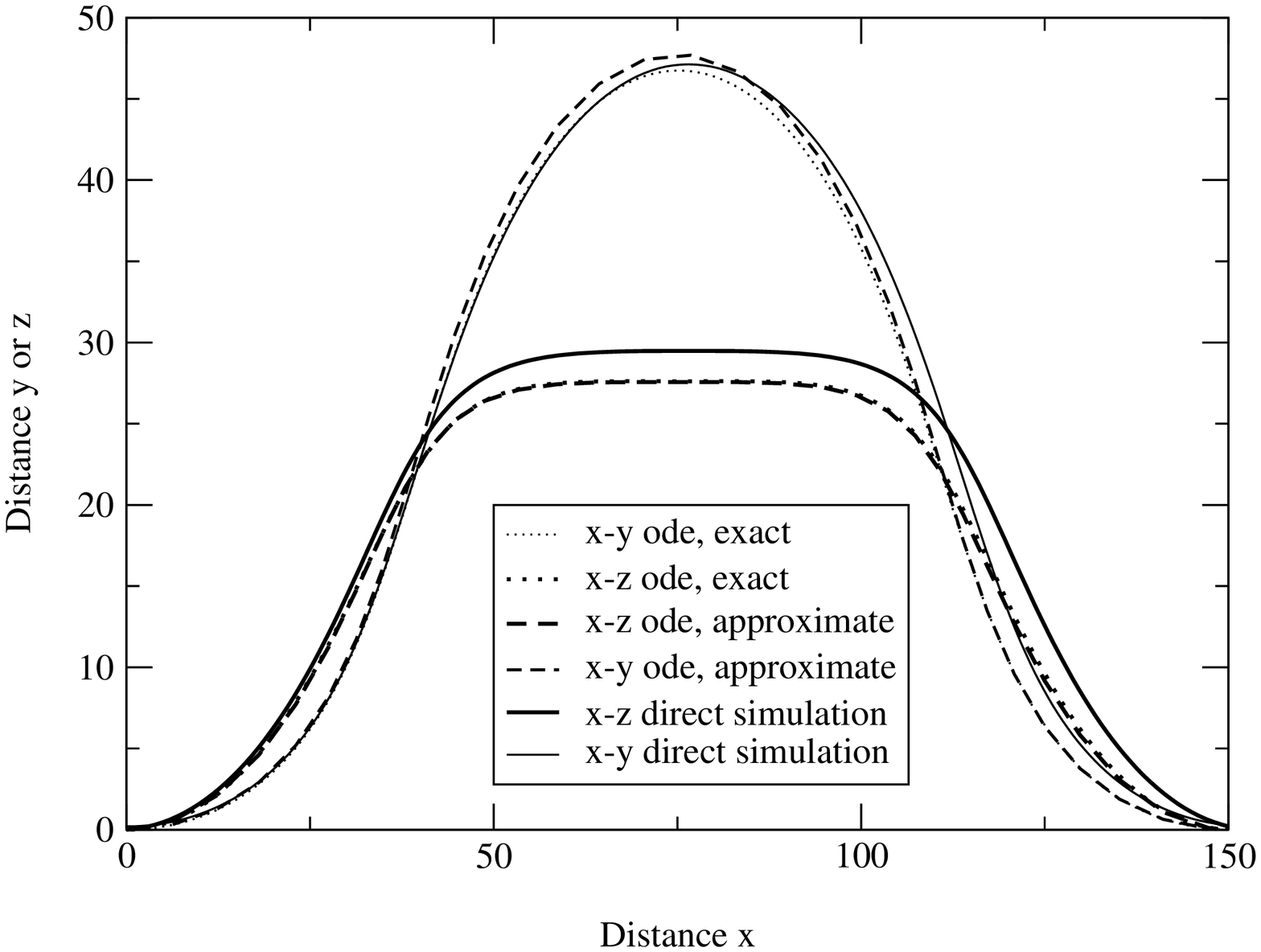}\\ (b)
\epsfxsize=4in\epsffile{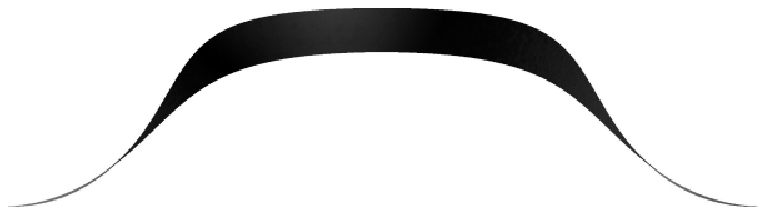} 
\caption{\small (a) Comparison of three separate methods to
  find shape of strip. The shape indicated as the exact solution
  results from solving Eq. \ref{eq:VAR4} with $u_4(0)=-9.26053$,
  $\dot\phi(0)=1.8129$, and $p_x=-4.4599$, leading to a solution with
  $\sqrt{g_0}=1.333$ and $\theta_f=3.1416$. The shape indicated as an
  approximate solution comes from evaluating Eq.~\ref{eq:vectors} for
  $\sqrt{g_{0}}=4/3$, $\theta_f=\pi$, and $R=63.7=200/\pi$. Finally,
  the shape resulting from direct simulation is obtained as a result
  of a direct numerical minimization of Eq.~\ref{eq:energy2}, for a
  strip 201 atoms long in the $x$ direction, and 12 layers wide in the
  $y$ direction, constrained to have horizontal period of 150 so that
  $\sqrt{g_{0}}=4/3$, and $R=63.7$. The energy predicted by the
  function Eq.~\ref{eq:dominant} with $L=200$ and $w=11\sqrt{3/4}$,
  and the approximate forms for $\dot\phi$ and $\dot\psi$ from Eq.
  \ref{eq:alpha} is $1.3571
  {\cal K}a^3$.  The corresponding calculation using exact results
  from Euler--Lagrange equations in Eq.~\ref{eq:VAR4} is $1.3557 {\cal
  K}a^3$ 
  The energy found from direct numerical minimization of
  Eq. \ref{eq:energy2} is $1.377 {\cal K}a^3$. (b)
  Three--dimensional visualization of the final shape resulting from
  the numerical minimization.
}

\label{fig:strip_comparison}
\end{center}
\end{figure}

Turning to Eq.~\ref{eq:vectors}, one finds that $\vec p$ must point
only in the $\hat x$ direction and
\begin{subequations}
\begin{eqnarray}
(C_2-C_1)\omega^2&=& p_x(1+\theta_0')\\
(C_1-C_2)(1-\alpha)\omega&=&p_x\theta_0'\omega'.
\end{eqnarray}
\end{subequations}
Using Eq.~\ref{eq:a_condition} to relate $\theta_0'$ and $\alpha$, and
Eq.~\ref{eq:pdefs}, one finds 
\begin{equation}
C_1=C_2
\label{eq:GOOD}
\end{equation}
or
\begin{equation}
\alpha=1+\omega^2.
\label{eq:BAD}
\end{equation}

Unfortunately, Eq.~\ref{eq:BAD} contradicts
Eq.~\ref{eq:a_value}. Therefore, the functional forms in
Eq.~\ref{eq:alpha} minimize the energy only if $C_1=C_2$, and
otherwise do not.

\section{Numerical Solutions}

\begin{figure}[!ht]
\epsfysize=3in\epsffile{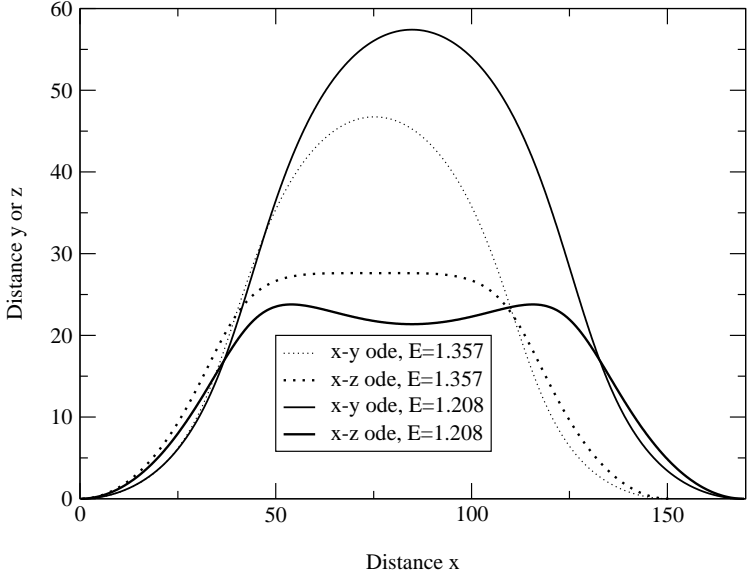}
\caption{\small Comparison of the lowest--energy shape from
  Fig. \ref{fig:strip_comparison2} with the lowest--energy shape found
  numerically by varying the initial conditions in Eqs.~\ref{eq:VAR4}
  while requiring $L/\lambda=\sqrt{g_0}=4/3$ and scaling the solutions by
  $R=200/\pi$.  The longer--wavelength solution has energy $1.2075
  {\cal K}a^3$. Lower--energy solutions probably exist, but my
  numerical routines have been unable to track beyond this point.} 
\label{fig:strip_comparison2}
\end{figure}

The Euler--Lagrange equations in Eq. \ref{eq:VAR4} constitute a closed set
of equations that determine all energy minima. It is surprisingly
tricky to employ them for this purpose, and I have only partial
results to report. Searching for periodic solutions of
Eqs.~\ref{eq:VAR4} proceeds in the following way: choose a value of
$p_x$, and the initial values of $u_4(\theta)$ and $\dot
\phi(\theta)$. Integrate forward in $\theta$ until reaching an angle
$\theta_f$ at which $\dot\psi(\theta_f)$ vanishes. It will not in
general be true that the solution at this point obeys
Eq.~\ref{eq:constraint}, or that $\hat r_1$ and $\hat r_2$ point along
the $\hat x$ and $\hat y$ axes as they must. However, by adjusting
$p_x$ alone, all of these boundary conditions can be achieved at
once. One still has the two initial values $\dot\phi(0)$ and $u_4(0)$
to vary. These allow one to vary $\theta_f$ and $\lambda$
independently. 

I have used a shooting algorithm to find solutions along these
lines. The process works, but is very delicate. Unless all constants
are chosen very near to their final values, the solutions of the
equations fly off into completely unphysical regions. The analytical
solutions of Section VIII are indispensable for the purpose of
locating good starting points for shooting solutions. Once a few
solutions have been located, additional solutions can be obtained by
using extrapolation to predict new values for initial conditions and
searching in a narrow range of parameter values around the
extrapolated predictions. 

When $C_1=C_2$ in Eq.~\ref{eq:MIN1}, the solution in Eq.
\ref{eq:vectors} does indeed minimize the energy when the wavelength
of the pattern is constrained to obey Eq.~\ref{eq:wavelength}.
However, even when $C_1=C_2$, the lowest--energy solution for a given
$\sqrt{g_0}$ is not given by demanding that the wavelength correspond
to this analytically tractable value. For a given
$\sqrt{g_0}=L/\lambda$, energies per length around 25\% lower are
obtained by allowing $\lambda$ and $L$ to increase; the energy
decreases monotonically as they become larger. The shooting procedure
eventually becomes unable to track additional solutions, and it is not
yet possible to say whether there are any energy minima to be found at
particular wavelengths. 
  
When $C_1=2$ and $C_2=3$ in Eq.~\ref{eq:MIN1}, as it should for an
isotropic two--dimensional material in accord with Eq.
\ref{eq:dominant}, the solution in Eq.  \ref{eq:vectors} provides a
good enough approximation to the true solution that the numerical
routines are able to converge by using its properties as a starting
point. The comments of the paragraph above otherwise do not need to be
changed.  As shown in Fig. \ref{fig:strip_comparison}, when boundary
conditions corresponding to a solution obeying Eq.~\ref{eq:wavelength}
are imposed upon the Euler--Lagrange equations, their solution is
almost indistinguishable from Eqs. \ref{eq:vectors}. However for the
same value of $\sqrt{g_0}$, the longer the value of the wavelength
$\lambda$, the lower the energy. There is no indication of an energy
minimum at a periodic 
solution. Fig. \ref{fig:strip_comparison2} compares the lowest--energy
solution I have found for $\sqrt{g_0}=4/3$ with the solution displayed
previously in Fig.~\ref{fig:strip_comparison}.

 \begin{figure}[!ht]
 \begin{center}
 \epsfxsize5in\epsffile{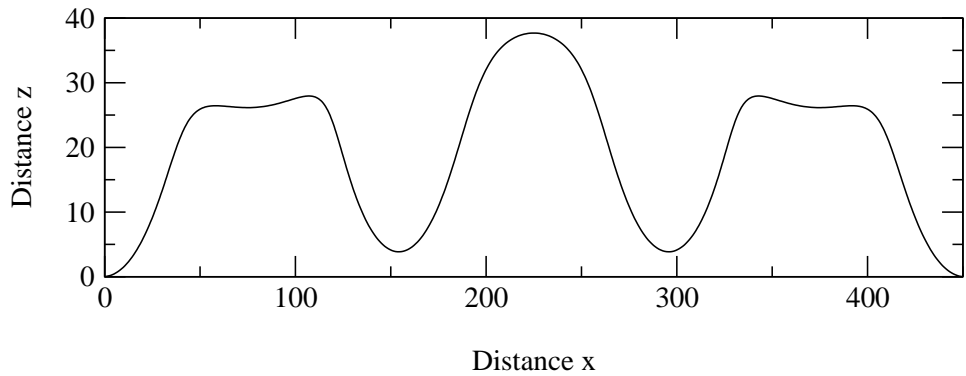}
 \caption{\small Numerical minimization of the same system described in
   Fig.~\ref{fig:strip_comparison}, but with $\lambda$ and $L$ three
   times as long. The total energy is $4.08 {\cal K}a^3$, which is less
   than the value $3\times 1.37 {\cal K}a^3$ one would obtain simply by
   pasting together three copies of the numerical solution in 
   Fig.~\ref{fig:strip_comparison}. This computation provides an
   additional hint that periodic solutions do not minimize the energy. }
 \label{fig:composite_strip}
 \end{center}
 \end{figure}
 An alternate way to check whether periodic solutions are favorable is
 to carry out direct numerical minimization in large systems.
 Returning to the numerical setting of
 Fig.~\ref{fig:strip_comparison}, I triple the length of the system,
 fix the boundaries at distance $3\lambda$, leave all else as before,
 and minimize the energy Eq.~\ref{eq:energy2}. The strip develops a
 composite structure in which the long wavelength is 3 times the short
 wavelength, and with an energy that is somewhat less than 3 times the
 energy of the structure 1/3 as long, as shown in
 Fig.~\ref{fig:composite_strip}.

Thus it is possible that the lowest--energy state for the simple metric in
Eq.~\ref{eq:linear_metric} involves a cascade of oscillations on many
scales, but the matter is not yet settled. Many other
problems remain to be addressed, including the application of the
ideas obtained here to the more complex metrics illustrated in
Fig.~\ref{fig:Membrane_setup}.

\acknowledgements
Eran Sharon discussed these problems with me for weeks, and spent
months carrying out experiments, before I finally began to see there
was something interesting to be done. Lorenzo Sadun helped us see what
could be gained from differential geometry. Thanks to Ralf Stephan for pointing
out Ref.~\cite{Nechaev.01}. Financial support from the National Science Foundation
(DMR-9877044 DMR-0101030) is gratefully acknowledged.

\bibstyle{Foundations_of_Physics}
\bibliography{membrane}

\begin{thebibliography}{11}
\expandafter\ifx\csname natexlab\endcsname\relax\def\natexlab#1{#1}\fi
\expandafter\ifx\csname bibnamefont\endcsname\relax
  \def\bibnamefont#1{#1}\fi
\expandafter\ifx\csname bibfnamefont\endcsname\relax
  \def\bibfnamefont#1{#1}\fi
\expandafter\ifx\csname citenamefont\endcsname\relax
  \def\citenamefont#1{#1}\fi
\expandafter\ifx\csname url\endcsname\relax
  \def\url#1{\texttt{#1}}\fi
\expandafter\ifx\csname urlprefix\endcsname\relax\def\urlprefix{URL }\fi
\providecommand{\bibinfo}[2]{#2}
\providecommand{\eprint}[2][]{\url{#2}}

\bibitem[{\citenamefont{Sharon et~al.}(2002)\citenamefont{Sharon, Roman,
  Marder, Shin, and Swinney}}]{Sharon.02}
\bibinfo{author}{\bibfnamefont{E.}~\bibnamefont{Sharon}},
  \bibinfo{author}{\bibfnamefont{B.}~\bibnamefont{Roman}},
  \bibinfo{author}{\bibfnamefont{M.}~\bibnamefont{Marder}},
  \bibinfo{author}{\bibfnamefont{G.-S.} \bibnamefont{Shin}}, \bibnamefont{and}
  \bibinfo{author}{\bibfnamefont{H.~L.} \bibnamefont{Swinney}},
  \bibinfo{journal}{Nature} \textbf{\bibinfo{volume}{419}},
  \bibinfo{pages}{579} (\bibinfo{year}{2002}).

\bibitem[{\citenamefont{Byrne et~al.}(2001)\citenamefont{Byrne, Timmermans,
  Kidner, and Martinssen}}]{Byrne.01}
\bibinfo{author}{\bibfnamefont{M.}~\bibnamefont{Byrne}},
  \bibinfo{author}{\bibfnamefont{M.}~\bibnamefont{Timmermans}},
  \bibinfo{author}{\bibfnamefont{C.}~\bibnamefont{Kidner}}, \bibnamefont{and}
  \bibinfo{author}{\bibfnamefont{R.}~\bibnamefont{Martinssen}},
  \bibinfo{journal}{Current Opinion in Plant Biology}
  \textbf{\bibinfo{volume}{4}}(\bibinfo{number}{1}), \bibinfo{pages}{38}
  (\bibinfo{year}{2001}).

\bibitem[{\citenamefont{Green et~al.}(1996)\citenamefont{Green, Steele, and
  Rennich}}]{Green.96}
\bibinfo{author}{\bibfnamefont{P.}~\bibnamefont{Green}},
  \bibinfo{author}{\bibfnamefont{C.}~\bibnamefont{Steele}}, \bibnamefont{and}
  \bibinfo{author}{\bibfnamefont{S.~C.} \bibnamefont{Rennich}},
  \bibinfo{journal}{Annals of Botany} \textbf{\bibinfo{volume}{77}},
  \bibinfo{pages}{515} (\bibinfo{year}{1996}).

\bibitem[{\citenamefont{Green}(1999)}]{Green.99}
\bibinfo{author}{\bibfnamefont{P.~B.} \bibnamefont{Green}},
  \bibinfo{journal}{American Journal of Botany} \textbf{\bibinfo{volume}{86}},
  \bibinfo{pages}{1059} (\bibinfo{year}{1999}).

\bibitem[{\citenamefont{Nechaev and Voituriez}(2001)}]{Nechaev.01}
\bibinfo{author}{\bibfnamefont{S.}~\bibnamefont{Nechaev}} \bibnamefont{and}
  \bibinfo{author}{\bibfnamefont{R.}~\bibnamefont{Voituriez}},
  \bibinfo{journal}{Journal of Physics A} \textbf{\bibinfo{volume}{34}},
  \bibinfo{pages}{11069 } (\bibinfo{year}{2001}).

\bibitem[{\citenamefont{Henderson and Taimina}(2002)}]{Taimina.02}
\bibinfo{author}{\bibfnamefont{D.}~\bibnamefont{Henderson}} \bibnamefont{and}
  \bibinfo{author}{\bibfnamefont{D.}~\bibnamefont{Taimina}},
  \bibinfo{journal}{Mathematical Intelligencer} \textbf{\bibinfo{volume}{23}},
  \bibinfo{pages}{17} (\bibinfo{year}{2002}).

\bibitem[{\citenamefont{Pogorelov}(1956)}]{Pogorelov.56}
\bibinfo{author}{\bibfnamefont{A.~V.} \bibnamefont{Pogorelov}},
  \emph{\bibinfo{title}{Differential Geometry}} (\bibinfo{publisher}{P
  Noordhoff N. V.}, \bibinfo{address}{Groningen}, \bibinfo{year}{1956}).

\bibitem[{\citenamefont{Landau and Lifshitz}(1986)}]{Landau.86}
\bibinfo{author}{\bibfnamefont{L.~D.} \bibnamefont{Landau}} \bibnamefont{and}
  \bibinfo{author}{\bibfnamefont{E.~M.} \bibnamefont{Lifshitz}},
  \emph{\bibinfo{title}{Theory of Elasticity}} (\bibinfo{publisher}{Pergamon
  Press}, \bibinfo{address}{Oxford}, \bibinfo{year}{1986}),
  \bibinfo{edition}{3rd} ed.

\bibitem[{\citenamefont{Mansfield}(1964)}]{Mansfield.64}
\bibinfo{author}{\bibfnamefont{E.~H.} \bibnamefont{Mansfield}},
  \emph{\bibinfo{title}{The Bending and Stretching of Plates}}
  (\bibinfo{publisher}{Pergamon}, \bibinfo{address}{New York},
  \bibinfo{year}{1964}).

\bibitem[{\citenamefont{Seung and Nelson}(1988)}]{Seung.88}
\bibinfo{author}{\bibfnamefont{H.~S.} \bibnamefont{Seung}} \bibnamefont{and}
  \bibinfo{author}{\bibfnamefont{D.~R.} \bibnamefont{Nelson}},
  \bibinfo{journal}{Physical Review A} \textbf{\bibinfo{volume}{38}},
  \bibinfo{pages}{1005} (\bibinfo{year}{1988}).

\bibitem[{\citenamefont{Lobkovsky and Witten}(1997)}]{Lobkovsky.97}
\bibinfo{author}{\bibfnamefont{A.~E.} \bibnamefont{Lobkovsky}}
  \bibnamefont{and} \bibinfo{author}{\bibfnamefont{T.~A.}
  \bibnamefont{Witten}}, \bibinfo{journal}{Physical Review E}
  \textbf{\bibinfo{volume}{55}}, \bibinfo{pages}{1577} (\bibinfo{year}{1997}).

\end{thebibliography}

\end{document}